%% file: main.tex
\documentclass[
    twocolumn,
	prd,
	amssymb,
	preprintnumbers,
	secnumarabic,
	nofootinbib]{revtex4-1}

\pdfoutput=1

\usepackage{graphicx}
\usepackage{enumitem}
\usepackage{latexsym}
\usepackage{amsfonts}
\usepackage{amssymb}
\usepackage{color}
\usepackage{amsmath}
\usepackage{slashed}
\usepackage{dcolumn}
\usepackage{verbatim}
\usepackage{float}
\usepackage{multirow}
\usepackage{xspace}
\usepackage[normalem]{ulem}
\usepackage[
pdfauthor={Nirmal Raj}]{hyperref}

\input{universalnewcommands.tex}

\newcommand{\DLensSource}{D_{\rm LS}}
\newcommand{\DLens}{D_{\rm L}}
\newcommand{\DSource}{D_{\rm S}}
\newcommand{\uT}{u_{1.34}}
\newcommand{\thetaE}{\theta_{\rm E}}
\newcommand{\rE}{r_{\rm E}}
\newcommand{\tE}{t_{\rm E}}
\newcommand{\vE}{v_{\rm E}}
\newcommand{\rholens}{\rho_{\rm lens}}

\newcommand{\rmax}{r_{90}}
\newcommand{\tmax}{t_{\rm m}}

\allowdisplaybreaks

\begin{document}

\title{Gravitational microlensing by dark matter in extended structures}
\author{Djuna Croon} \email{dcroon@triumf.ca}
\author{David McKeen}
\email{mckeen@triumf.ca}
\author{Nirmal Raj} \email{nraj@triumf.ca}
\affiliation{TRIUMF, 4004 Wesbrook Mall, Vancouver, BC V6T 2A3, Canada}

\date{\today}

\begin{abstract}
Dark matter may be in the form of non-baryonic structures such as compact subhalos and boson stars. 
Structures weighing between asteroid and solar masses may be discovered via gravitational microlensing, an astronomical probe that has in the past helped constrain the population of primordial black holes and baryonic \acro{MACHO}s. 
We investigate the non-trivial effect of the size of and density distribution within these structures on the microlensing signal, and constrain their populations using the \acro{EROS}-\osn{2} and \acro{OGLE-IV} surveys. 
Structures larger than a solar radius are generally constrained more weakly than point-like lenses, but stronger constraints may be obtained for structures with mass distributions that give rise to caustic crossings or produce larger magnifications.
\end{abstract}

\maketitle

\section{Introduction}

All our evidence for  dark matter is gravitational.
Its presence is seen in 
its gravitational pull on stars,
gravitational lensing of light from distant galaxies,
and gravitational imprints on both the thermal fluctuations of photons from the early universe 
and the structure of the universe on scales galactic and higher.  
While extensive attempts are at large to produce dark matter at colliders and to detect ambient dark matter directly or, in the remnants of its annihilation, indirectly,
an appealing possibility is to unmask its microscopic identity in yet more gravitational phenomena.
One such opportunity is provided by the recent inauguration of gravitational wave astronomy~\cite{Bertone:2019irm},
another by seeking anomalies in pulsar timing arrays~\cite{PulsarTiming}, 
and yet another by the observation of capture of dark matter sped up by the steep gravitational potentials of compact stars~\cite{GoldmanNussinov,DKHNS}.
Astronomical efforts have also been long underway to constrain black holes or faint astronomical objects (``\acro{MACHO}s") as dark matter candidates via the technique of gravitational microlensing, the observation of temporary, all-wavelength brightening of a background star whose light is deflected by the transiting object. 
In this paper, we study the microlensing signals of dark matter in macroscopic structures made of non-standard states, i.e. structures that are {\em not} \acro{macho}s. 

\begin{figure}
    \centering
     \includegraphics[width=0.48\textwidth]{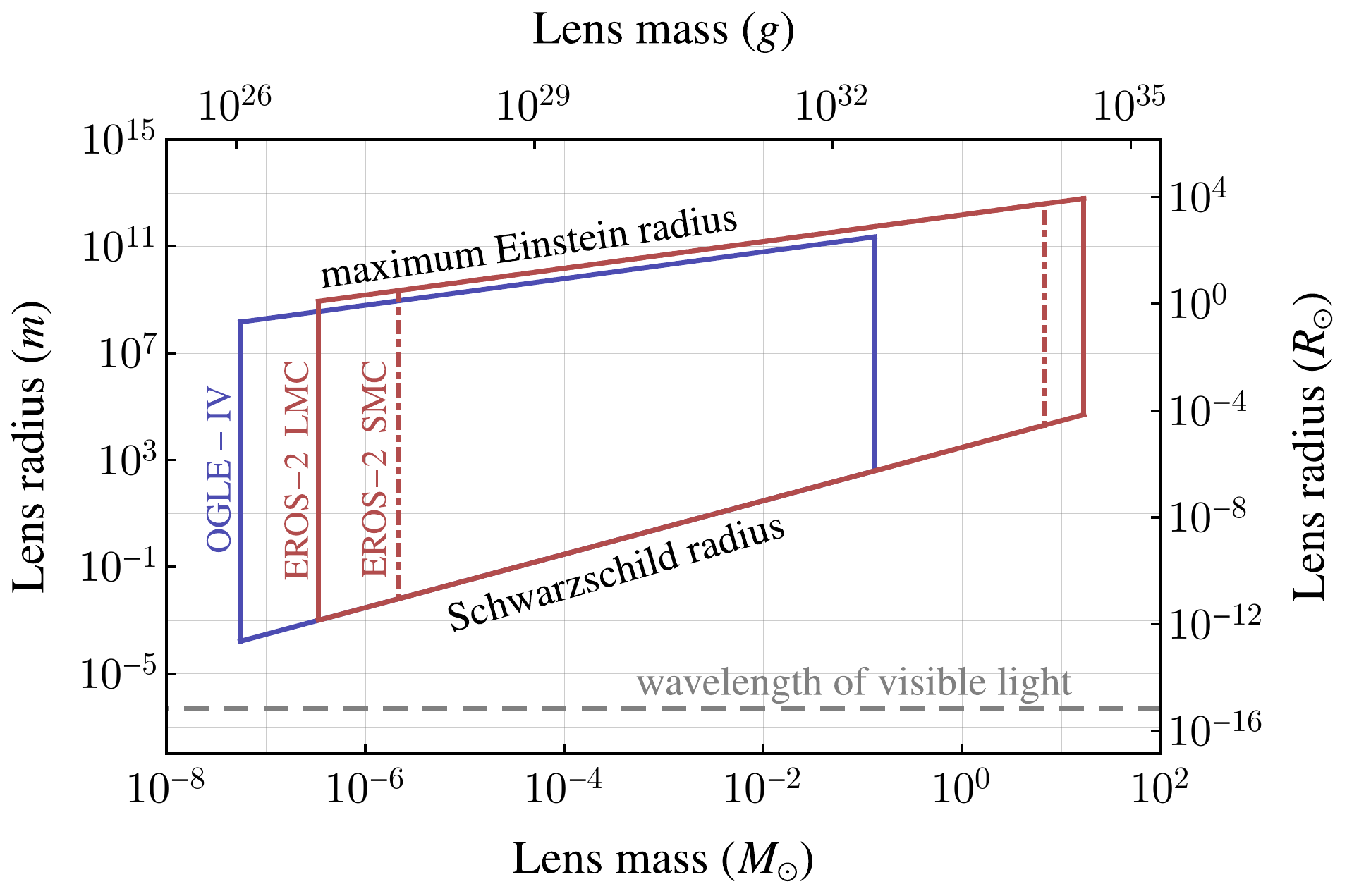}
    \caption{A heuristic estimate of the masses and sizes of dark matter structures that can be probed  by the \acro{eros}-\osn{2} and \acro{ogle-iv} microlensing surveys of the Magellanic Clouds and Milky Way Bulge respectively.
    Here constant detection efficiencies and zero backgrounds are assumed for simplicity. 
    For a fixed mass, the smallest lens size corresponds to the Schwarzschild radius of a black hole with the corresponding mass; the magnification is suppressed for lenses much smaller than the wavelength of light used in a survey. 
    The largest lens size for a given mass that can be probed by the survey in question is approximately the maximum Einstein radius at that mass, above which the lens becomes too diffuse to magnify source stars efficiently.
    The lowest and highest masses probed by a survey are determined by the observational cadences of the survey.
    See Sec.~\ref{sec:limits} for further details.
     }
    \label{fig:RvsM}
\end{figure}

The key distance scale in microlensing is the Einstein radius, the distance of closest approach of light rays seen by an observer when the source, lens, and observer lie along a line, as the light bends around a lens of mass $M$.
Up to $\mathcal{O}$(1) factors, the Einstein radius is the geometric mean of the distance to the source star and the Schwarschild radius associated with the lens, $2G M/c^2$. 
Microlensing surveys are typically sensitive to stars that are 10--1000 kpc away, and to transit times of minutes to years. Thus, for galactic dark matter with speed dispersion $\sim 10^{-3}~c$, microlensing is sensitive to dark matter masses ranging from asteroid to solar masses.
Several scenarios predict a large population of dark matter structures in this mass range. 
Non-standard cosmologies such as an early period of matter domination or vector boson production during inflation may enhance the growth of small-scale density perturbations, as a result of which most of the dark matter may survive currently in compact subhalos~\cite{ErickcekKris,Barenboim:2013gya,Fan2014,CoDecay,VectorDMInflation}.  
In a similar vein, axion dark matter can form dense ``miniclusters"~\cite{Fairbairn:2017dmf}.
The use of weak and strong gravitational lensing to probe such sub-structure has been already discussed~\cite{weakstronglensing}.
Dark matter may also lurk in quantum structures such as boson stars, which are kept from collapsing under self-gravity by kinetic pressure or self-repulsion (e.g.~\cite{Ruffini:1969qy,Gleiser:1988rq,BSReview,Croon:2018ybs}).
An observation of the characteristic microlensing signals of any of these structures would have enormous implications for cosmology, star formation, and galaxy evolution.

Gravitational microlensing depends crucially on the spatial extent of the lens.
Dark matter structures much smaller than the microlensing Einsten radius imitate point-like lenses (such as black holes and \acro{macho}s), and those much greater would barely lens the source star due to the diffuse spread of their mass.
But for structures with characteristic size comparable to the Einstein radius
the signal is non-trivial, and therefore so are the sensitivities of microlensing surveys to the population of dark matter in these structures.
The signal also depends sensitively on how the net mass of a lens is distributed within it, as this determines how it bends spacetime and deflects light.
Earlier studies have dealt with microlensing by dense primordial hydrogen-helium gas clouds~\cite{WidrowHClouds},
and by axion miniclusters~\cite{Fairbairn:2017dmf,Blinov_2020}.
In this work we extend these studies to dark matter subhalos with various density profiles, as well as to boson stars, and obtain constraints on their populations from the \acro{eros}-\osn{2}~\cite{EROS2} and \acro{ogle-iv}~\cite{OGLEIV5yrII} surveys.
In Fig.~\ref{fig:RvsM} we display, in the space of lens size vs. lens mass, the best-case sensitivities of these surveys to these extended structures. 
In Section~\ref{sec:basics} we explain how to determine these sensitivities.
A recent microlensing survey by the Subaru telescope~\cite{Subaru} also probes our scenario, however this survey was sensitive to a range of Einstein radii so small that the finite angular extent of source stars affects the signal non-trivially.
In forthcoming work~\cite{finitesq} we investigate the effect of finite-sized sources on microlensing by finite-sized lenses.

This paper is laid out as follows.
In Section~\ref{sec:basics} we review the basics of gravitational microlensing, describing the treatment of extended lenses.
In Section~\ref{sec:limits} we estimate the rate of microlensing events and derive constraints on the fraction of dark matter in non-baryonic structures.
In Section~\ref{sec:concs} we outline future avenues of research and conclude. 
In the appendices we collect detailed derivations of lens-plane-projected mass profiles of the various structures we consider, and provide a method for rapid numerical computations of event rates.

\begin{figure}
    \centering
     \includegraphics[width=0.48\textwidth]{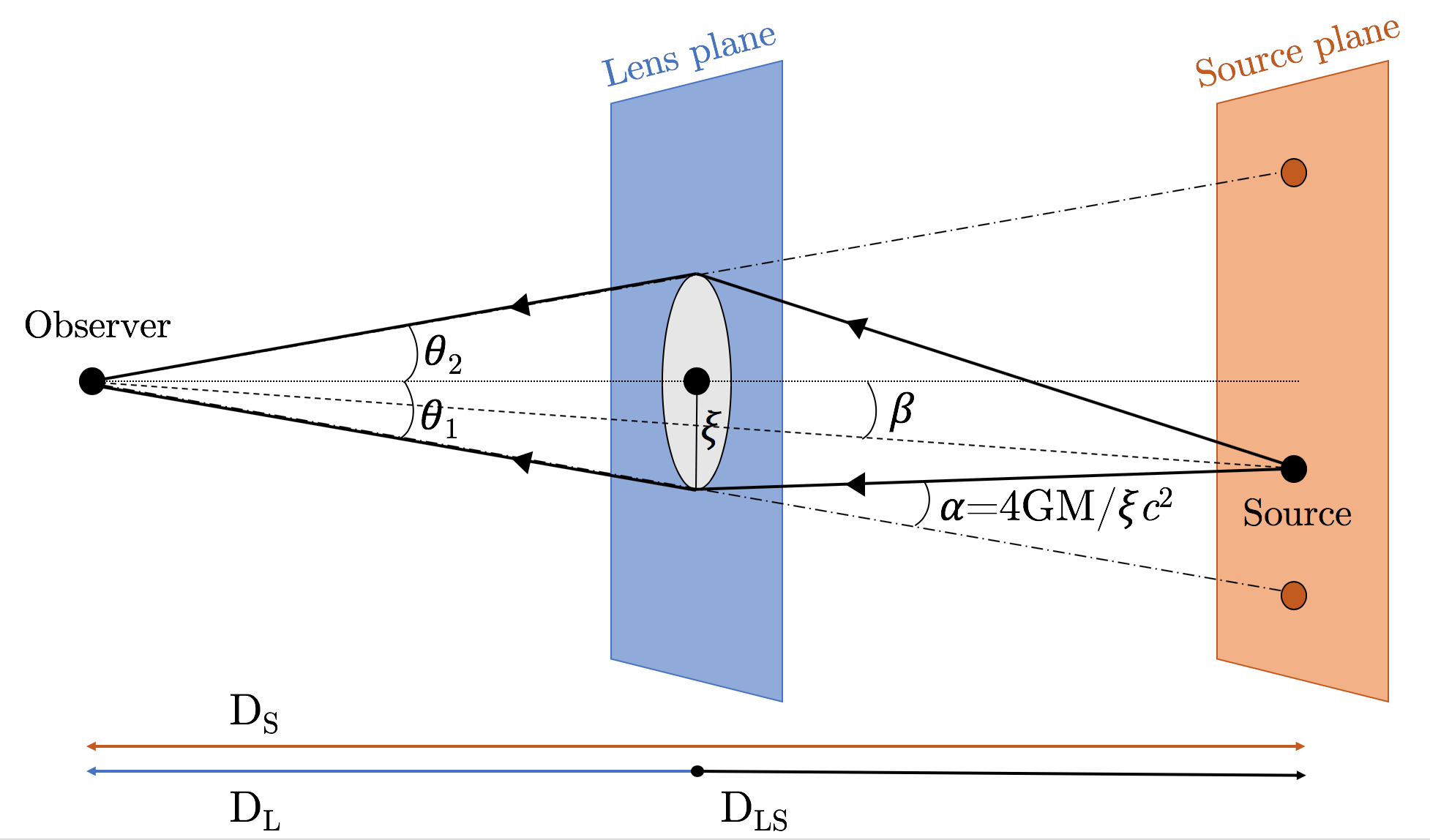}
    \caption{
    The geometry of the microlensing setup for a point-like lens, seen to produce two images after deflection. 
   From this geometry follows the lensing equation [Eq.~\eqref{eq:lens}] governing the path of light rays, which determines the magnification [Eq.~\eqref{eq:mag}].
   }
    \label{fig:geometry}
\end{figure}

\section{Microlensing of Extended Objects}
\label{sec:basics}

\begin{figure}
    \centering
         \includegraphics[width=0.48\textwidth]{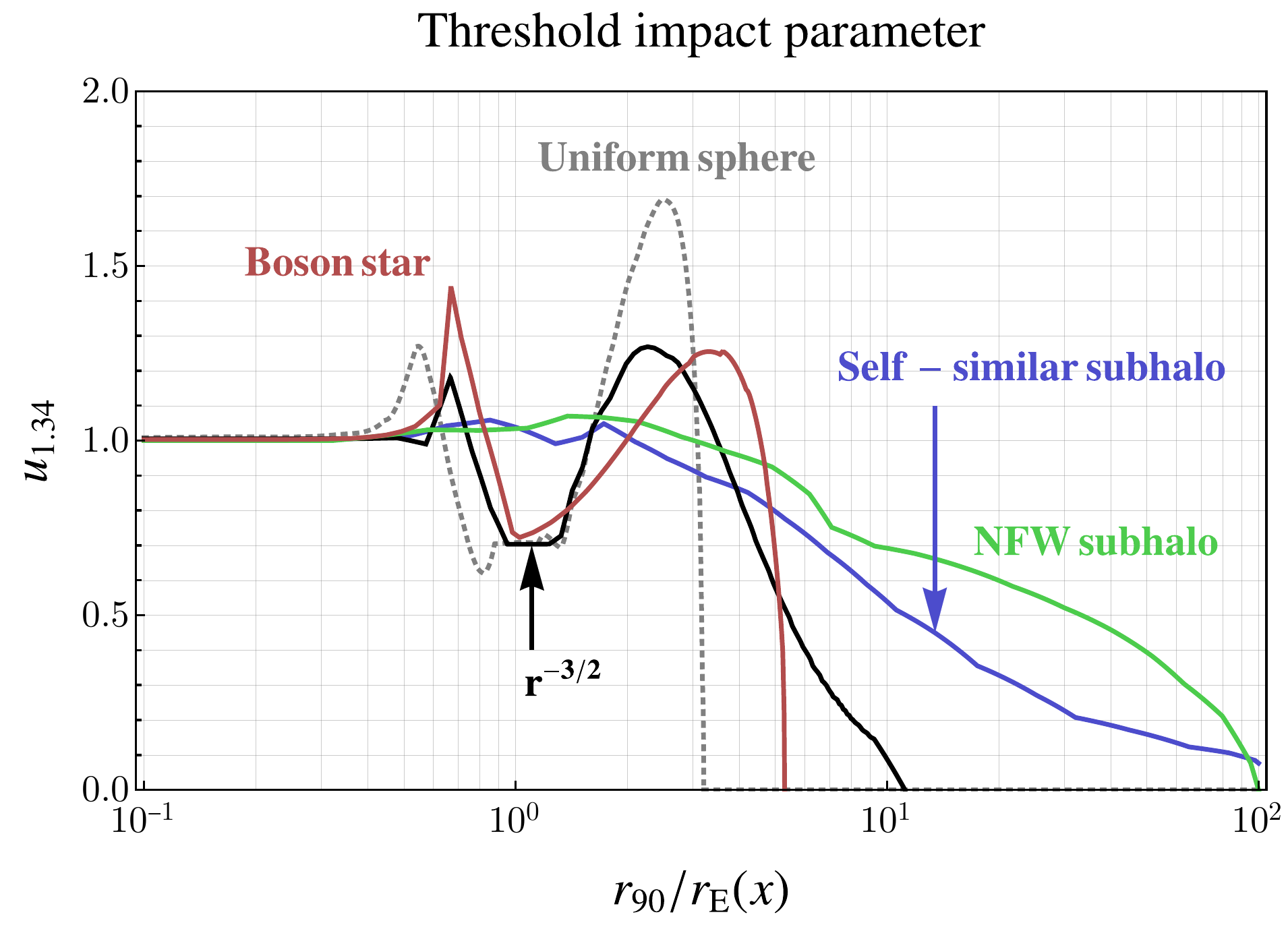}
      \includegraphics[width=0.48\textwidth]{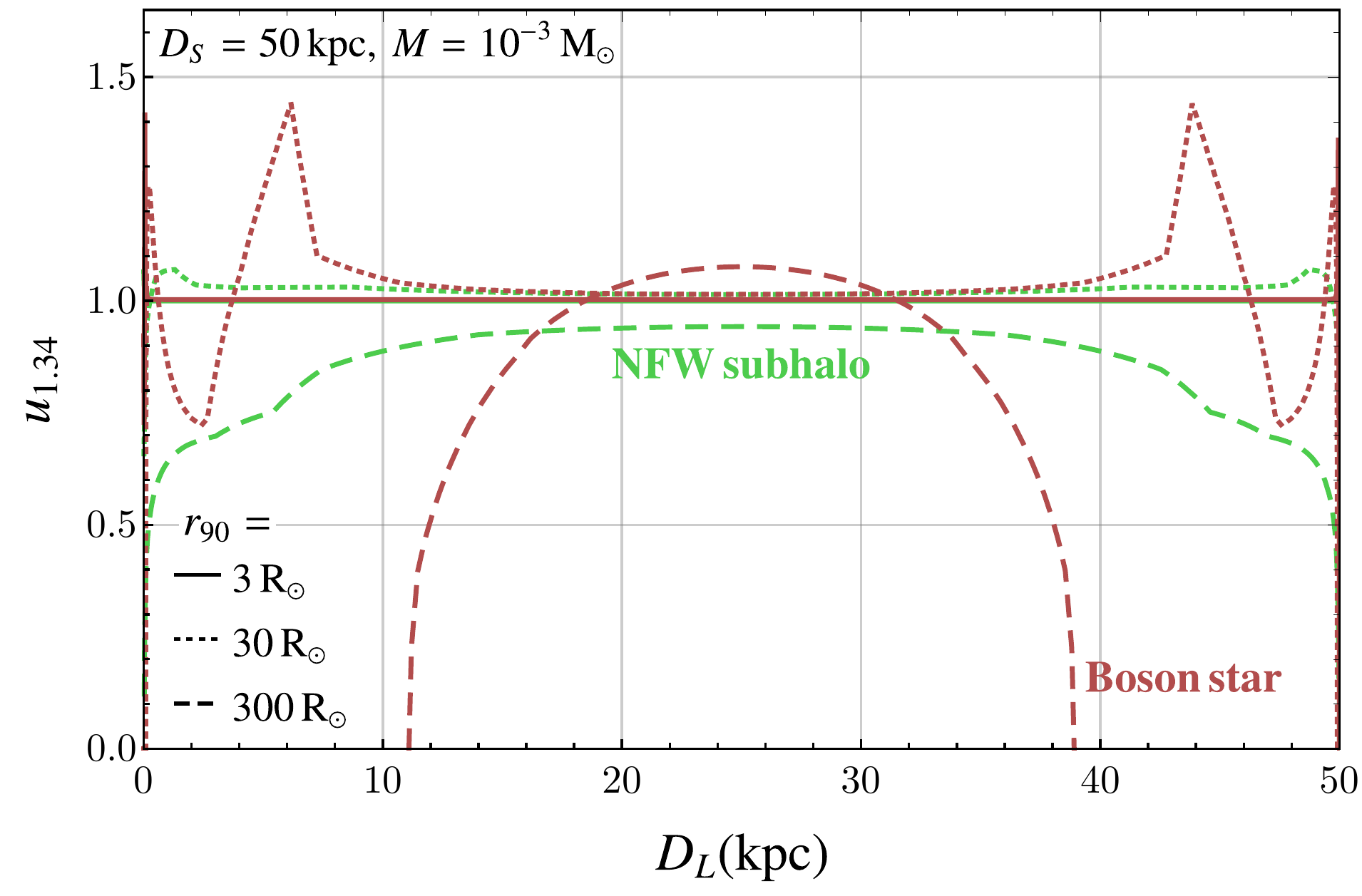}
    \caption{
     Threshold impact parameter in units of the point-like Einstein radius, as defined in Eq.~\eqref{eq:uTdef}, for various lens species. 
    {\bf \em Top}: as a function of the lens radius containing 90\% of the total mass $r_{90}$, in units of the point-like Einstein radius. 
    {\bf \em Bottom}: as a function of the distance to the lens $\DLens$ for \acro{NFW} subhalos and boson stars of various sizes with total mass $M=10^{-3}M_\odot$ and distance to the source $\DSource=50~\rm kpc$.
    In both panels, the spiky features appear for lens sizes and distances corresponding to the lens crossing a caustic, where the magnification formally diverges.
    The threshold impact parameter defines the cross-sectional radius of the ``detector volume" at a microlensing survey, transits across which are counted as events. Hence the information in these plots is key to obtaining limits on dark matter populations, as done in Sec.~\ref{sec:limits}.
    }
    \label{fig:uT}
\end{figure}

In this section, we review some basics of gravitational microlensing, largely     following the treatment in Ref.~\cite{Narayan:1996ba}.

Figure~\ref{fig:geometry} depicts the geometry of the setup. 
The observer-lens, lens-source, and observer-source distances are $\DLens$, $\DSource$, and $\DLensSource$ respectively; 
the lens center and the source (an image and the source) subtend an angle $\beta$ ($\theta$) at the observer: in general, more than one image may be formed.
Our calculations simplify as $\DLens$, $\DSource$, and $\DLensSource$ are much larger than all other scales in the problem,
resulting in small angular deflections of value $\alpha = 4 GM/(c^2 \xi)$ that only occur, in this approximation, when starlight encounters the ``lens plane'' perpendicular to the observer-source axis.  

Next we assume that the lens is spherically symmetric\footnote{For studies on microlensing by aspherical objects, see, e.g., Refs.~\cite{Bozza:2004va,Tsapras_2018}.} with density distribution $\rho(r)$ so that the total mass $M=4\pi\int_0^\infty dr r^2\rho(r)$. 
The lensing equation that determines the path of light rays after deflection may then be written as
\beq
\beta=\theta-\frac{\thetaE^2}{\theta}\frac{M(\theta)}{M}~,
\label{eq:lens}
\eeq
where 
\bea
\nn M(\theta)&=&2\pi \DLens^2\int_0^\theta d\theta^\prime \theta^\prime \Sigma(\theta^\prime)~,\\
\Sigma(\theta)&=&\int_{-\infty}^\infty dz\,\rho\left(\sqrt{\DLens^2\theta^2+z^2}\right)~,
\label{eq:Mthetasigmatheta}
\eea
the latter quantity being the surface mass density projected onto the lens plane.
The point-like Einstein angle $\thetaE$ is given by~\cite{Einstein:1956zz}
\begin{equation}
\thetaE \equiv\sqrt{\frac{4GM}{c^2}\frac{\DLensSource}{\DLens\DSource}}~,
\label{eq:thetabar}
\end{equation}
obtained as the value of $\theta$ for a point-like lens ($M(\theta) \ra M$) at $\beta = 0$.
This in turn defines the point-like Einstein radius\footnote{In this paper, Einstein angle and radius, $\thetaE$ and $\rE$, should be understood to refer only to the values corresponding to a point-like lens of mass $M$.} $\rE \equiv \DLens \thetaE$ on the lens plane.
The volume contained within circles of radius $\rE$ whose centers are along the line of sight is sometimes known as the ``lensing tube".
From Eq.~\eqref{eq:thetabar} it may seen that the lensing tube for point-like lens is an ellipsoid.

Given a lens position $\beta$, Eq.~\eqref{eq:lens} may be solved to determine the image position(s), $\theta$. 
While gravitational microlensing does not alter the {\em luminosity} of the source, the images subtend solid angles that are different from the source, proportionally altering the {\em flux} 
received.\footnote{One assumes here that the lens is transparent to light (as is true for structures of dark matter) lest it occult the image.
We thank Yue Zhao for raising this point on either side of the Pacific.
}
Thus the magnification induced by an image is the ratio of its angular extent to that of the source:
\beq
\mu (\theta (\beta))=\left|\frac{\theta}{\beta}\frac{d\theta}{d\beta}\right|.
\label{eq:mag}
\eeq
The lightcurve, or the magnification as a function of time $t$, for a lens with velocity $v$ and minimum impact parameter $\xi_{\rm min}$ is now determined by setting $\beta \rE = \sqrt{\xi^2_{\rm min} + v^2t^2}$. 
As the lens approaches and leaves the vicinity of the observer-source axis, its image brightens and dims, the hallmark signature of gravitational microlensing. 
Whether such an occurrence is actually observable depends on the minimum detectable magnification for a given telescope, as well as the range of cadences at the microlensing survey, which sets the transit timescales to which it is sensitive.

For a point-like lens at some impact parameter (in units of Einstein radius) 
$u \equiv \xi/\rE = \beta/\thetaE$, we get from Eqs.~\eqref{eq:lens} and \eqref{eq:mag} the total magnification
\begin{equation}
    \begin{aligned}
 \mu_{\rm tot} &= \frac{u^2+2}{u\sqrt{u^2+4}} \\
&\xrightarrow{u=1} 1.34~.
\end{aligned}
\label{eq:magpointlike}
\end{equation}

For a point-like lens, a microlensing event is defined as a transit across the lensing tube, which acts as the ``detector volume". 
That is, a transit is counted as an event if it produces a total magnification $\mu_{\rm tot}\geq \mu_{\rm T}$, where it is conventional to take $\mu_{\rm T} =$ 1.34. 
Following this convention, we will also use this minimum magnification of a transit to define events for extended lenses. 

Let us now characterize the \emph{microlensing efficiency of an extended lens} compared to that of a point-like lens. 
To do so, we define the quantity $\uT$ as the impact parameter for a lens such that all smaller impact parameters produce a magnification above the threshold:
\bea
\mu_{\rm tot}(u\le \uT) &\ge& 1.34~.
\label{eq:uTdef}
\eea

For a point-like lens, $\uT$ = 1.
For an extended lens, it is clear from Eqs.~\eqref{eq:lens} and \eqref{eq:mag} that $\uT$ depends on its mass profile, $m(\theta) \equiv M(\theta)/M$.
In Appendix~\ref{app:massprofiles} we derive these mass profiles for all the lens species we consider: 
\begin{itemize}
    
 \item {\bf self-similar subhalos}, products of the isolated gravitational collapse of primordial density perturbations~\cite{Bertschinger:1985pd}. 
 Their density profiles scale as -9/4 powers of the radius,  
    
    \item lenses with density profiles that scale as {\bf -3/2 powers of the radius}. These are inspired by the inner profiles of ultra-compact minihalos (\acro{ucmh}s), said to form at redshifts $\geq 1000$ in regions where overdensities are very large, $\delta \rho/\rho \gsim 10^{-3}$~\cite{UCMHProfile,*Delos:2018ueo},
    
    \item {\bf Navarro-Frenk-White (NFW) subhalos}, suggested to be the products of hierarchical clustering triggered by, {\em e.g.} cold dark matter~\cite{NFW},
    
    \item {\bf boson stars}, gravitationally stable structures composed of scalar fields~\cite{BSReview},
    
    \item {\bf uniform spheres} (of constant density) as a toy model.
\end{itemize}

In the top panel of Fig.~\ref{fig:uT} we plot the various $\uT$ as a function of $\rmax/\rE(x)$, where $\rmax$ is the radius within which 90\% of the lens mass is contained, and $x \equiv \DLens/\DSource$.
As expected, $\uT \ra 1$ ($\uT \ra 0$) for $\rmax \ll \rE$ ($\rmax \gg \rE$), while the most interesting features arise in the intermediary regime.
The spikes in $\uT$ at $\rmax \lsim \rE$ for the \acro{ucmh}-like subhalo, boson star and uniform sphere are caused by the lens crossing a ``caustic", an impact parameter at which the number of images changes discontinuously and produces infinite magnification (in reality regulated by the finite extent of the source).
For these lens species, regions where $\uT < 1$ correspond to there being only one image contributing to $\mu_{\rm tot}$ at $u = \uT$, however even a single image can have large $\mu_{\rm tot}$ for $\rmax \gsim \rE$, so that $\uT > 1$ in this region.
For \acro{nfw} and self-similar subhalos there are no caustics, and $\uT > 1$ for $\rmax \gsim \rE$. 
Their mass distributions make them efficient lenses even at very large $\rmax$, so that $\uT$ decreases gradually as $r_{\rm max}/\rE$ is increased. 
The above information on microlensing efficiency is key to obtaining microlensing event rates and constraints on lens populations, a task to which we will turn in the next section.

Let us illustrate the above features with concrete examples. 
In the bottom panel of Figure~\ref{fig:uT} we have plotted $\uT$ versus $\DLens$ for \acro{NFW} subhalos and boson stars of mass $10^{-3} M_\odot$ for various $\rmax$: $3 R_\odot, 30 R_\odot, 300 R_\odot$, where $R_\odot = 6.96 \times 10^9$~ is the solar radius.
The distance to the source is here assumed to be 50 kpc, corresponding to the distance to the Large Magellanic Cloud.
The maximum $\rE$ in this setup is 70 $R_\odot$.
Lenses much smaller than this have $\uT \simeq 1$ everywhere, viz., behave like point-like lenses.
The 30~$R_\odot$ (300~$R_\odot$) \acro{nfw} subhalo is slightly more (less) efficient than point-like lenses everywhere in the lensing tube.
Both the 30~$R_\odot$ and 300~$R_\odot$ boson stars are more efficient than point-like lenses in the middle of the lensing tube, and less efficient near the source and observer. 
The 30~$R_\odot$ boson star is also seen to spike in efficiency about $5$~kpc off the source or observer, which occurs due to it crossing a caustic at these distances.

\begin{figure*}
    \centering
     \includegraphics[width=0.46\textwidth]{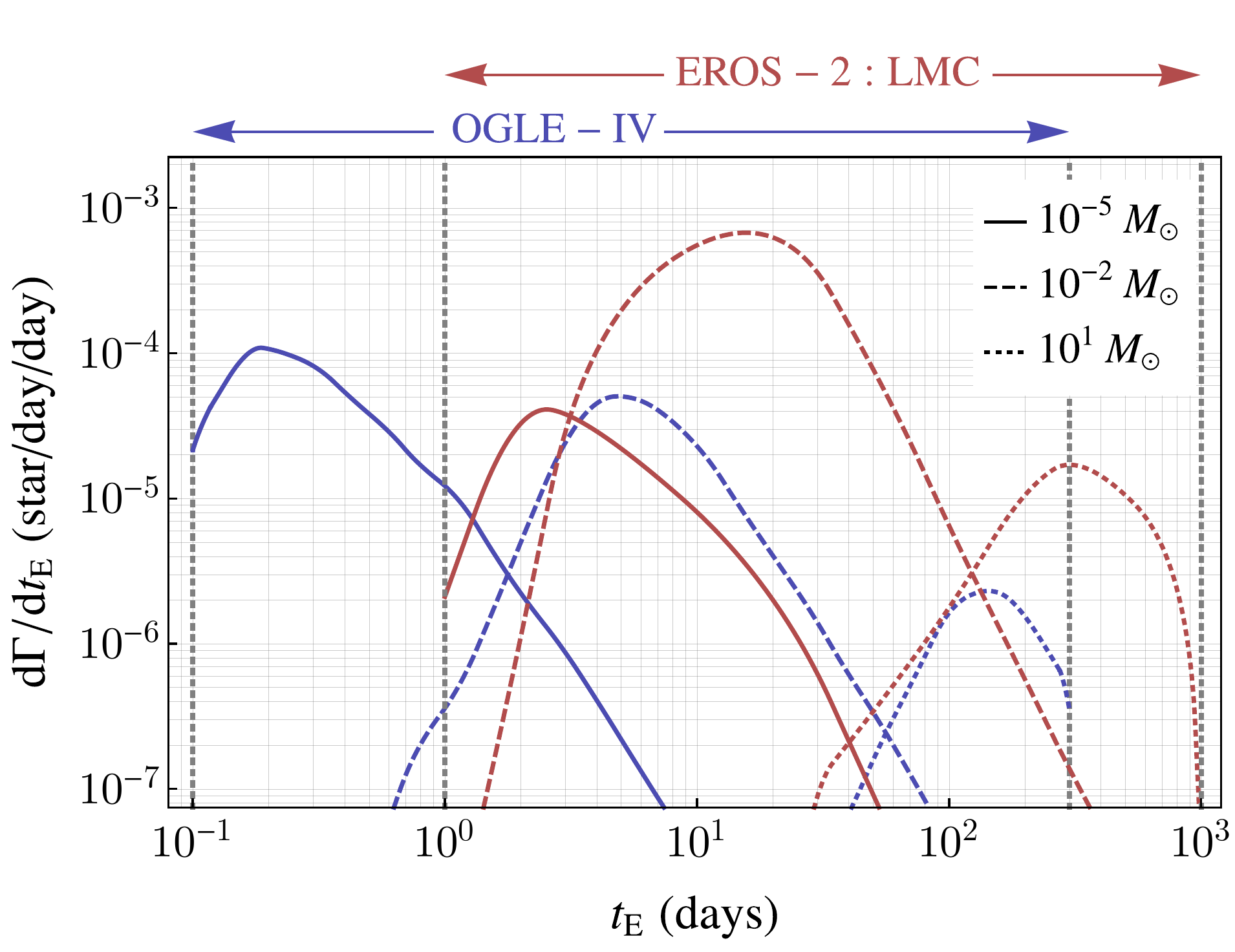} 
     \includegraphics[width=0.45\textwidth]{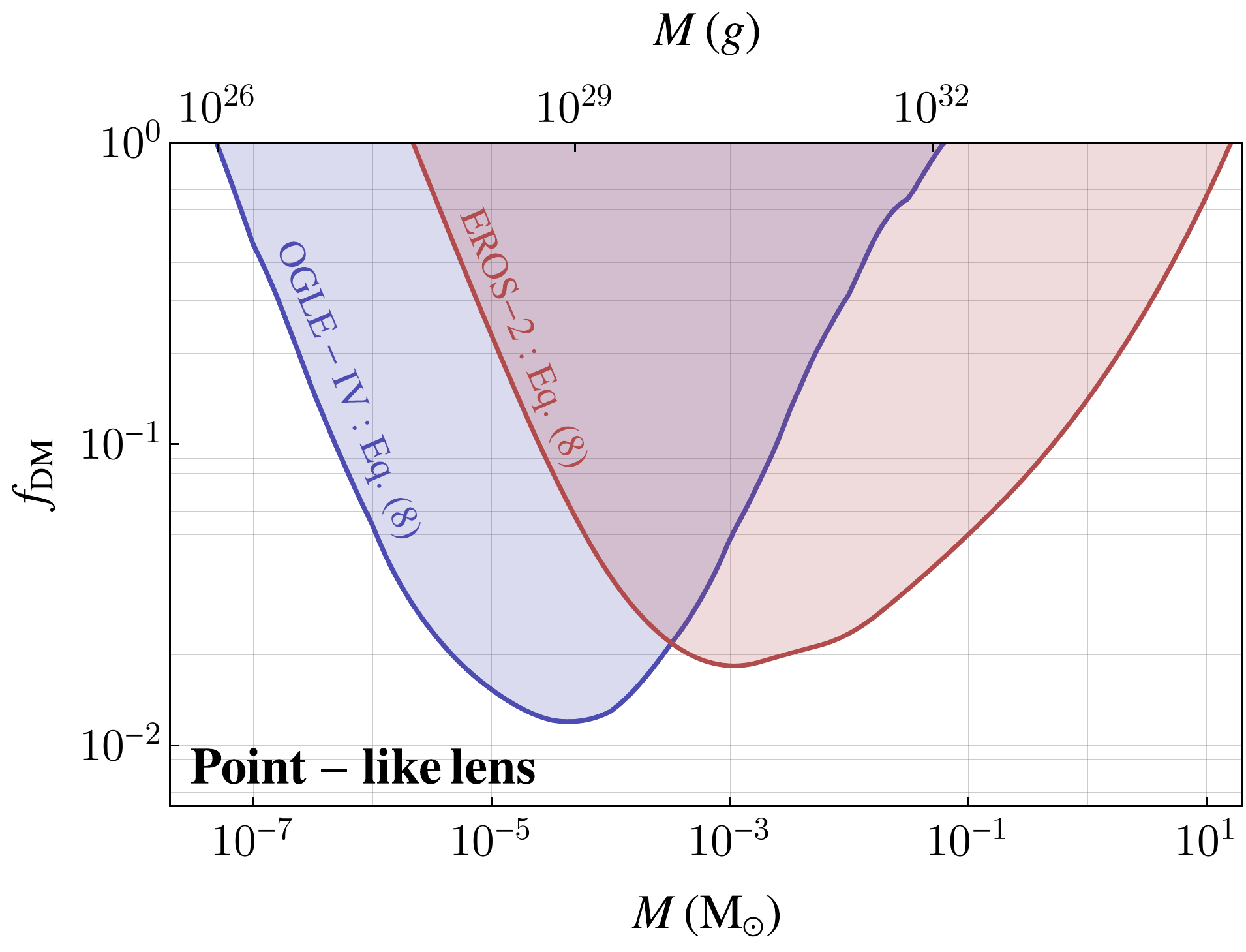}
    \caption{{{\bf \em Left}: Differential event detection rates per unit exposure of dark matter in point-like lenses at the \acro{eros}-\osn{2} and \acro{ogle-iv} microlensing surveys, as obtained from Eq.~\eqref{eq:dGammadxdt} and Table~\ref{tab:surveyparams}. 
    The low-$\tE$ (high-$\tE$) contribution of very light (very heavy) lenses to the event rates are left out due to limitations of the cadences.
    {\bf \em Right}: 90\% \acro{c.l.} limits on the fraction of point-like lenses making up the dark matter density, as estimated in Sec.~\ref{sec:limits}. 
  See text for further details.}
    }
    \label{fig:dNdtE}
\end{figure*}

\begin{figure*}
    \centering
    \includegraphics[width=0.99\textwidth]{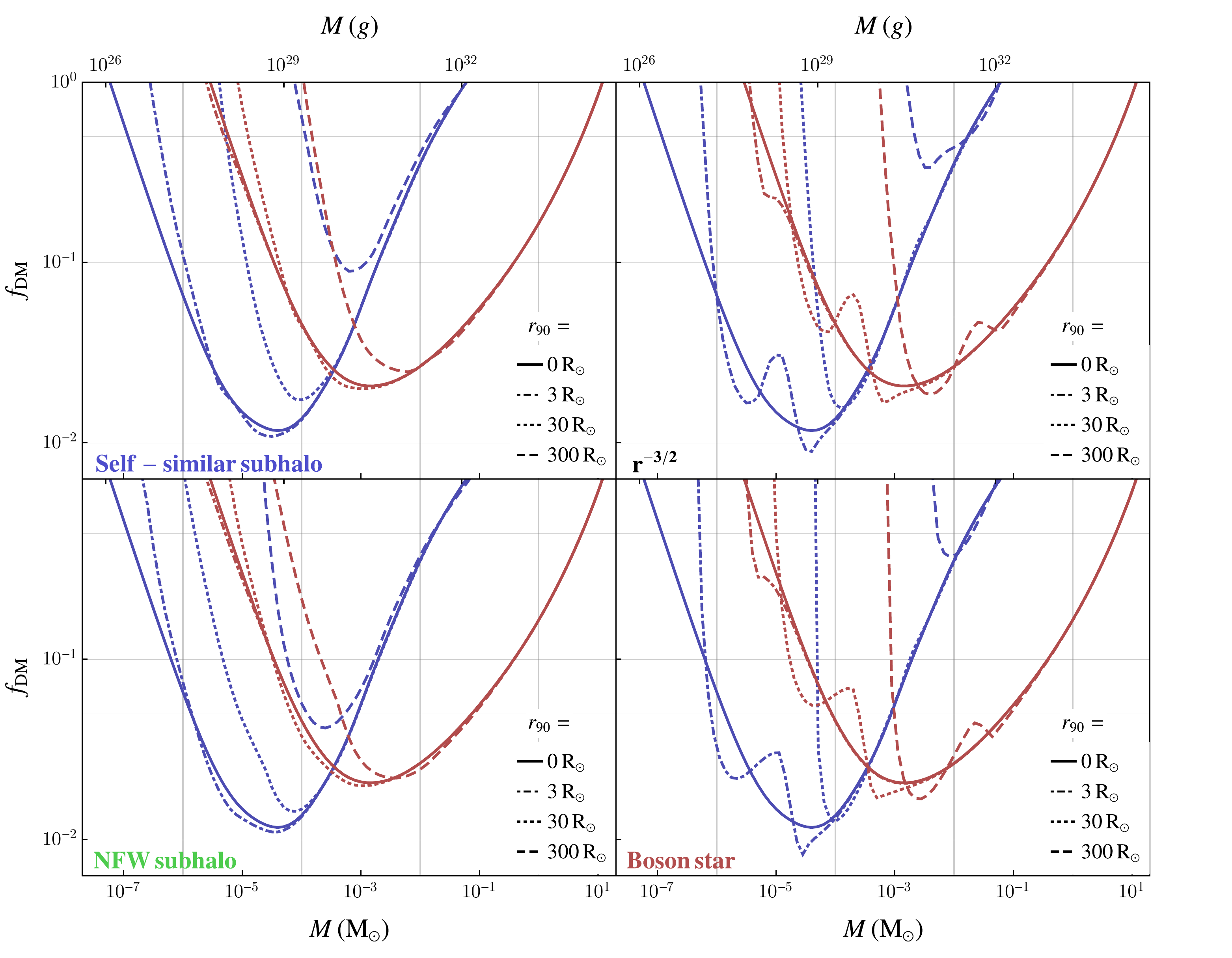} 
    \caption{
   90\% \acro{c.l.} limits from the microlensing surveys \acro{EROS}-\osn{2} (red curves) and \acro{OGLE-IV} (blue curves) on the fraction of dark matter in non-baryonic extended structures, as obtained in Sec.~\ref{sec:limits}. 
    Structures smaller than {$10~R_\odot$ ($R_\odot$)} may be approximated as a point-like lens at \acro{EROS}-\osn{2} (\acro{OGLE-IV}), resulting in constraints essentially the same as in Fig.~\ref{fig:dNdtE}.
    Larger structures magnify background stars non-trivially depending on their size and internal mass distribution (see Fig.~\ref{fig:uT}), resulting in structure-specific constraints.
    }
    \label{fig:$$}
\end{figure*}

\section{Event Rates and Constraints}
\label{sec:limits}

\begin{table*}
\begin{center}
\begin{tabular}{ l  l  l  l l l } 
\hline
survey & source field & ($\DSource$, $\ell$, $b$)  & $N_\star$ & $T_{\rm obs}$ (day) & $t_{\rm E}$ range (day) \\
\hline
\hline
\multirow{2}{6em}{\acro{EROS}-\osn{2}~\cite{EROS2}} & \acro{LMC} & (50 kpc, 280.46$^\circ$, -32.89$^\circ$) & 5.49 $\times 10^6$ & \multirow{2}{4em}{2500} & \multirow{2}{4em}{[1,1000]}  \\ 
                                       & \acro{SMC} & (60 kpc, 302.81$^\circ$, -44.33$^\circ$) & 0.86 $\times 10^6$ & \\ 
                                      \hline
\acro{OGLE-IV~\cite{OGLEIV5yrI,OGLEIV5yrII,Niikura:2019kqi}} & \acro{MW} Bulge & (8.5 kpc, 1.09$^\circ$, -2.39$^\circ$) & 4.88 $\times 10^7$ & 1826 &  [0.1,300] \\
\hline
\end{tabular}
\end{center}
\caption{Survey parameters used for placing constraints in Section~\ref{sec:limits}. 
The third panel provides the location of the source in terms of (distance, longitude, latitude), 
{taken from Ref.~\cite{SIMBADDatabase}.}
}
\label{tab:surveyparams}
\end{table*}

Having defined a microlensing event in Sec.~\ref{sec:basics}, we now estimate the rate of events collected in microlensing surveys. 
Here we follow the treatment in Ref.~\cite{Griest:1990vu}, which takes into account the distribution of dark matter velocities, assumed to be Maxwell-Boltzmann, as well as the density of dark matter along the line of sight to the source, $\rholens(x) = f_{\rm DM} \rho_{\rm DM}(x)$, where $f_{\rm DM}$ is the mass fraction of lenses making up the dark matter density $\rho_{\rm DM}$.
For a singe source star and unit exposure time, the differential event rate with respect to $x$ and event timescale $t_{\rm E}$ is then obtained as 
\beq
\frac{d^2\Gamma}{dx d\tE} = \varepsilon(\tE) \frac{2\DSource}{v_0^2 M} f_{\rm DM} \rho_{\rm DM}(x) \vE^4(x) e^{-\vE^2(x)/v^2_0}~,
\label{eq:dGammadxdt}
\eeq
where $\vE(x) \equiv 2 \uT(x)\rE(x)/\tE$ with $\rE(x)$ given in Eq.~\eqref{eq:thetabar} 
and $\uT(x)$ plotted in Fig.~\ref{fig:uT}, 
$v_0 = 220$~km/s is the dark matter circular speed in the galaxy, 
and $\varepsilon(\tE)$ is the efficiency of telescopic detection.
Equation~\eqref{eq:dGammadxdt} assumes that the source and the observer are static in the microlensing setup, which yields event rates to within 10\% accuracy~\cite{Niikura:2019kqi}.
It also assumes that all lenses in a population have a single mass $M$. Generalizing to other mass distributions is straightforward.
The total number of events is then given simply by
\beq
N_{\rm events} = N_\star T_{\rm obs} \int^1_0 dx \int^{t_{\rm E, max}}_{t_{\rm E, min}} d\tE \frac{d^2\Gamma}{dx d\tE}~,
\label{eq:Nevents}
\eeq
where 
$N_\star$ is the number of source stars used in the survey,
$T_{\rm obs}$ is the total observation time, 
and $t_{\rm E, min}$ ($t_{\rm E, max}$) is the minimum (maximum) timescale of an event in the survey.
In Appendix~\ref{app:speed} we briefly describe how to numerically evaluate the above integral rapidly.

The parameters used for \acro{eros}-\osn{2} and \acro{ogle-iv} are given in Table~\ref{tab:surveyparams}.
For evaluating $N_{\rm events}$ in Eq.~\eqref{eq:Nevents}, we use the locations of the source stars and the detection efficiencies provided in Refs.~\cite{EROS2,Niikura:2019kqi}, and assume an isothermal profile for the Milky Way halo,
\bea
\nn \rho_{\rm DM} (r) &=& \frac{\rho_{\rm s}}{1+(r/r_{\rm s})^2}~,\\
r &\equiv& \sqrt{R^2_{\rm Sol} - 2 x R_{\rm Sol} \DSource \cos \ell \cos b + x^2 \DSource^2}~,
\eea
where $R_{\rm Sol}$ = 8.5~kpc, $\rho_{\rm s} = 1.39$~GeV/cm$^3$, and $r_{\rm s} = 4.38$~kpc are the galactic radial distance of the Sun, the core density, and core radius respectively~\cite{PPPPCookbook}; 
$\ell$ and $b$ are the longitude and latitude of the source in galactic co-ordinates.
Picking an isothermal galactic halo, in which $v_0$ does not vary with distance, simplifies our calculation of the event rate; we have checked that our results are robust against the choice of the halo profile.

In Fig.~\ref{fig:RvsM} we show a heuristic estimate of lens sizes and masses probed by the \acro{eros}-\osn{2} surveys of either Magellanic Cloud, and by the \acro{ogle-iv} survey of the Milky Way Bulge, assuming zero foreground, and a constant detection efficiency of 
$\varepsilon(\tE) = 0.2$ ($\varepsilon(\tE) = 0.05$) and event timescales 
\{$t_{\rm E, min}$, $t_{\rm E, max}$\}~=~\{1 d, 500 d\} (\{0.6 d, 10 d\}) for \acro{eros}-\osn{2} (\acro{ogle-iv}), an approximation that qualitatively characterizes the surveys in Refs.~\cite{EROS2,Niikura:2019kqi}.
Lenses cannot be more compact than black holes, hence their sizes are bounded from below by their Schwarzschild radius $\propto M$.
Moreover, the microlensing geometric setup breaks down for lens sizes much smaller than the photonic wavelength spectrum of the telescope, since the effects of wave optics greatly suppress the magnification~\cite{waveoptics}.
In Fig.~\ref{fig:RvsM} we have indicated this with a horizontal dashed line.
 As argued in Section~\ref{sec:basics}, lens sizes that can be constrained by microlensing are approximately bounded from above by their point-like Einstein radius $\propto M^{1/2}$.
 
At a given microlensing survey, there is a minimum ($t_{\rm E, min}$) and maximum ($t_{\rm E, max}$) transit time, set respectively by the maximum and minimum observational cadences.
For small $M$, we have $\vE \gg 1$, and evaluating Eq.~\eqref{eq:Nevents} in this limit, $N_{\rm events} \propto M/t^3_{\rm E, min}$. 
Thus the detection rate is suppressed as we decrease $M$: these lenses transit too briefly even for the survey's highest cadence. 
For large $M$, we have $\vE \ll 1$, and in this limit $N_{\rm events} \propto t^3_{\rm E, max}/M^2$. 
Here the detection rates are suppressed as we increase $M$: these lenses are too scarce in number, and transit too long for the survey's lowest cadence.
In Fig.~\ref{fig:RvsM}, $N_{\rm events} > 1$ for the mass range within the vertical lines corresponding to a survey, and $N_{\rm events} < 1$ outside.

To inspect the above limitations in more detail, in Fig.~\ref{fig:dNdtE} left-hand panel we show for point-like lenses ($\uT = 1$) the distance-integrated event rates $d\Gamma/d\tE$ at \acro{eros}-\osn{2} and \acro{ogle-iv}.
We see that these surveys sample most transits arising from point-like lenses of masses around $10^{-5} M_\odot$.
However, low-$\tE$ transits of lenses that are much lighter are missed.
This is because lighter lenses generate thinner lensing tubes, across which high-velocity transits from the tail of the Maxwell-Boltzmann distribution last for very short times, below the cadence thresholds of the telescope.
High-$\tE$ transits of lenses that are very heavy are also missed.
This is because these lenses generate very thick lensing tubes, across which low-velocity transits, albeit magnifying some background star, may not {\em alter} the source flux appreciably over the maximum timescale of transit to which the telescope is sensitive.

\subsection{Limits on point-like lenses}

The features discussed above result in the 90\%~\acro{c.l.} limits plotted in the right-hand panel of Fig.~\ref{fig:dNdtE}, on the fraction of dark matter in point-like lenses as a function of their mass.
To obtain the \acro{EROS}-\osn{2} limits, we used Eq.~\eqref{eq:Nevents} and set $N_{\rm events} = 3.9$, corresponding to the number of events expected at the 90\% \acro{c.l.} for the one event observed, assuming Poisson statistics.
Our limits differ somewhat from those estimated by the \acro{EROS} collaboration, who count events as~\cite{EROS2}
\beq
N^{\rm EROS}_{\rm events} = N_\star T_{\rm obs} \frac{2}{\pi} \frac{\varepsilon(\tE)}{\langle\tE\rangle}~\tau~,
\label{eq:EROScount}
\eeq
where $\tau$ is the ``optical depth", the probability of finding a lens in the lensing tube at any given instant:
\beq
\tau =  \DSource \int_0^1 dx \ \frac{f_{\rm DM} \ \rho_{\rm DM}(x)}{M} \pi \rE^2(x)~.
\eeq

In the mass range constrained by \acro{EROS}-\osn{2} there are also weak bounds from the \acro{macho} collaboration, which do not change the limits appreciably when combined with \acro{EROS}-\osn{2}~\cite{EROSMACHOCombined}.
Weaker limits also exist from the \acro{EROS}-\osn{1} survey~\cite{EROS1}; these would strengthen our limits at most by 20\% for large $M$, and are completely superseded by \acro{ogle-iv} for small $M$.
Finally, the observation of the blue supergiant \acro{MACS J}\osn{1149} \acro{LS}\osn{1} at a distance of 4.3 Gpc, interpreted as a point-like lens crossing a caustic and producing $\mathcal{O}$(10$^3$) magnification, provides a constraint in this mass range~\cite{CausticCrossingBound}.
However, we do not display it here as the astrophysics is relatively uncertain and the \acro{eros}-\osn{2} limit could almost entirely cover the constrained region (see Fig.~8 of Ref.~\cite{CausticCrossingBound}).

To obtain the \acro{OGLE-IV} limits one must account for $\mathcal{O}$(1000) events observed in their 5-year dataset, which agree at the 1\% level with astrophysical models of standard foreground events~\cite{Niikura:2019kqi}.\footnote{We note the recent appearance of \acro{OGLE-IV}'s 8-year dataset~\cite{OGLEIV8yr} containing 5790 observed events as opposed to 2622 in the 5-year dataset.
As the standard foreground has not been estimated for the new data (a task beyond our scope), we use the 5-year dataset for obtaining limits, remarking that we do not expect the limits to improve markedly due to increased foregrounds.} 
There are also six events near $\tE \sim$ 0.1 days for which there is no satisfactory explanation~\cite{OGLEIV5yrII,Niikura:2019kqi,Scholtz:2019csj},
but here we will adopt the null hypothesis that they constitute the foreground.
Using Ref.~\cite{Niikura:2019kqi}'s binning of events in $\tE$, for every bin $i$ we define $N^{\rm DM}_i$ as the number of dark matter-induced events obtained from Eq.~\eqref{eq:Nevents} and $N^{\rm FG}_i$ as the foreground count.
Then defining $N^{\rm SIG}_i \equiv N^{\rm FG}_i + N^{\rm DM}_i$, we use the quantity~\cite{PDGStats}
\beq
\kappa = 2 \sum_{i = 1}^{N_{\rm bins}} \left[ N^{\rm FG}_i - N^{\rm SIG}_i + N^{\rm SIG}_i \ln \frac{N^{\rm SIG}_i}{N^{\rm FG}_i} \right]
\eeq
and obtain the 90\% \acro{c.l.} Poissonian limit by locating ($f_{\rm DM}, M$) for which $\kappa$ = 4.61. 
Our resultant limit on the right-hand panel of Fig.~\ref{fig:dNdtE} is in good agreement with that obtained in Ref.~\cite{Niikura:2019kqi}.

\subsection{Limits on extended lenses: main results}

Next we obtain constraints for extended lenses of various sizes and density profiles by setting $\uT$ in Eq.~\eqref{eq:dGammadxdt} to the values plotted in Fig.~\ref{fig:uT}.
We display these limits in Fig.~\ref{fig:$$}, which are the main results of our paper.
For $\rmax \lsim 10~R_\odot$ ($\rmax \lsim R_\odot$), the \acro{eros}-\osn{2} (\acro{ogle-iv}) limits are the same as that for point-like lenses seen in Fig.~\ref{fig:dNdtE}.
For $\rmax$  in these ranges, the lenses are smaller than the smallest $\rE$ to which each survey is sensitive, hence $\uT \ra 1$ as seen in Fig.~\ref{fig:uT}.
As we increase $\rmax$, the limits for all lens species generally weaken for small $M$, where $\rE$ is small so that $\uT < 1$ for $\rmax/\rE > 1$ as seen in Fig.~\ref{fig:uT}.
In particular, the \acro{ogle-iv} limits for $\rmax = 30 R_\odot$ are $M \lsim 10^{-6}$--$10^{-5}~M_\odot$, weakening by 1--2 orders of magnitude with respect to point-like lenses.
The limit weakens further by $\mathcal{O}(100)$ for $\rmax = 300 R_\odot$.
Similarly, the \acro{eros}-\osn{2} limit on $M$ weakens by 1--2 orders of magnitude with respect to point-like lenses for $\rmax = 300 R_\odot$.

For the $r^{-3/2}$ and the boson star profiles, however, 
we notice additional features: for small $M$ there are two regions where the limits on $f_{\rm DM}$ are weaker than point-like lenses, and two where they are stronger.
We can understand this from the behavior of their $\uT$ in Fig.~\ref{fig:uT}; for fixed $\rmax$, scanning from left to right on this plot roughly corresponds to scanning from right to left on Fig.~\ref{fig:$$}. 
Thus, (1) for large $M$ the lens is point-like ($\uT \ra$ 1); 
(2) as we lower $M$ we enter the region where the lens is efficient ({\em i.e.} $\uT >$ 1) due to caustic crossings, giving stronger limits;
(3) as we lower $M$ further, the lens is inefficient ({\em i.e.} $\uT <$ 1) due to the contribution of a sole point-like image to the magnification, giving weaker limits;
(4) as we lower $M$ even further, the lens is efficient again ({\em i.e.} $\uT >$ 1) due to large magnification from one image, giving stronger limits again;
(5) finally, for very small $M$ the lens is too spatially diffuse to magnify the source efficiently ({\em i.e.} $\uT \ll$ 1), giving weaker limits again.

We end this section by remarking that we have not used any information about the lightcurves of events in deriving our constraints. 
For some lens species, e.g. boson stars, extra features in the lightcurves that may arise from caustic crossings could be used to better distinguish them from foregrounds, potentially giving improved limits or sensitivity.

\section{Discussion}
\label{sec:concs}

In this work, we estimated constraints from the gravitational microlensing surveys \acro{eros}-\osn{2} of the Magellanic Clouds and \acro{ogle-iv} of the Galactic Bulge,
on the population of dark matter in non-baryonic structures of 
self-similar subhalos,
density profiles that scale like the inner regions of ultra-compact minihalos,
\acro{NFW} subhalos,
and boson stars.
Our main results are summarized in Fig.~\ref{fig:$$}.
In deriving these limits we assumed that the source stars were point-like, an approximation that breaks down for the high-cadence survey of M31 by Subaru that is sensitive to small Einstein radii~\cite{Subaru}.
Computing the magnification of finite-sized stars by finite-sized lenses and estimating the ensuing limits on dark matter populations is the subject of our forthcoming work~\cite{finitesq}.

Our work is applicable to several avenues of research.
While we have estimated population limits on four representative extended structures, they may also be estimated for subhalos of other density profiles~\cite{WidrowHClouds,PulsarTiming}, 10 AU-sized dark stars~\cite{DarkStarsFreese}, $R_\odot$-sized mirror stars~\cite{MirrorStarsToronto,*MirrorStarsToronto2}, 
primordial dark matter halos~\cite{Savastano_2019},
extended structures formed by mirror dark matter~\cite{Cline2020}, and so on. 
Microlensing surveys of the Galaxy by the space-based \acro{wfirst}~\cite{WFirstProjection} and Euclid~\cite{EuclidProjection} (the Earth-based \acro{lsst}~\cite{LSSTProjection}) would probe sub-Earth mass (stellar mass) dark matter structures. 
The stellar mass range can also be probed by lensing of Type-Ia supernovae~\cite{SNeLensing}. 
Using parallax measurements in lensing can probe asteroid to planet masses~\cite{Parallax,ParallaxL2}.
Should numerous microlensing events be detected, the data on lightcurves may help us estimate the subhalo mass function and perform dark matter astrometry, in particular help identify departures from the Standard Halo Model such as the presence of tidal streams, as already constrained by weak lensing in the time domain~\cite{HalometryNYU,HalometryNYU2}.
Finally, it would be interesting to investigate the effect on our constraints of varying the magnification threshold in Eq.~\eqref{eq:uTdef}.

On the whole, microlensing is a promising gravitational probe to discover dark matter structures bred by novel cosmologies and astrophysics.



\appendix

\section{Mass profiles and lensing equations}
\label{app:massprofiles}

In this appendix we give some more information about the mass profiles $M(\theta)/M$ appearing in the lensing equation in Eq.~\eqref{eq:lens} for various lens species, discuss features of the resultant solutions to the lensing equation, and obtain expressions for the magnification.
These aspects show up as features in the threshold impact parameter $\uT$ (Fig.~\ref{fig:uT}), which are relevant for counting microlensing events and placing constraints.

Let us begin with generalities.
Expressing angles in units of the Einstein angle $\thetaE$ or, equivalently, distances on the lens plane in units of the Einstein radius ($u \equiv \beta/\thetaE= \DLens \beta/\rE$, $t \equiv \theta/\thetaE = \DLens \theta/\rE$) allows us to rewrite Eqs.~\eqref{eq:lens} and \eqref{eq:Mthetasigmatheta} as
\begin{equation}
u=t-\frac{m(t)}{t},
\label{eq:lens2}
\end{equation}
where $m(t) \equiv M( \thetaE t)/M$ describes the distribution of the lens mass projected onto the lens plane. For a spherically symmetric density profile $\rho(r)$,
\beq
m(t) = \frac{\int_0^{t} d\sigma \sigma \int^\infty_0 d\lambda\, \rho(\rE\sqrt{\sigma^2+\lambda^2})}{\int_0^\infty d\gamma \gamma^2 \rho(\rE\gamma)}~.
\label{eq:mtmaster}
\eeq

From Eq.~\eqref{eq:mag}, the magnification can be written as
\bea
\nn \mu &=& \left|\frac{t}{u}\frac{dt}{du}\right| \\
 &=& \left|1-\frac{m(t)}{t^2}\right|^{-1}\left|1+\frac{m(t)}{t^2}-\frac{1}{t}\frac{dm(t)}{dt}\right|^{-1}~.
\label{eq:mu2}
\eea

From this it is seen that the only way in which the total lens mass $M$ and the distances $\DLens, \DLensSource, \DSource$ enter the problem is through their contributions to $\rE$ in Eq.~\eqref{eq:thetabar}. 
At a fixed $\rE$, the density profile of the lens $\rho (r)$ turns up as $m(t)$ in Eq.~\eqref{eq:lens2}.
Solving the lensing equation, we can then use Eq.~(\ref{eq:mu2}) to determine the magnification of the image(s) as a function of $u$. 
We perform this calculation for specific lenses in the following subsections. In what follows, we will make some approximations that illustrate properties of the mass profile $m(t)$ and the resulting solutions of the lensing equation---however, to obtain the limits shown above, we have numerically calculated the $m(t)$ profiles without resorting to any approximate forms.

\subsection{Uniform sphere}
\label{app:subsec:sphunif}
As a warm up, we will study a spherical lens of uniform density. This distribution has the virtue of being analytically tractable and shares some qualitative features with other lenses such as the boson star profiles we will find below. The density of such an object can be written as $\rho(r) = \rho_0 \Theta(r_{\rm m} - r)$ where $r_{\rm m}$ is the radius of the sphere. From Eq.~\eqref{eq:mtmaster} we obtain
\begin{equation}
m(t)=\left\{
\begin{array}{ll}
1-\left(1-{t^2}/{t_{\rm m}^2}\right)^{3/2}, &  \left|t\right|<t_{\rm m}
\\
1~, &  \left|t\right|\geq t_{\rm m},
\end{array}
\right.
\label{eq:mtunif}
\end{equation}
where $\tmax \equiv r_{\rm m}/\rE$. The lensing equation~\eqref{eq:lens2} is now quintic in $t$, and depending on $b$ and $\tmax$, gives either one or three real solutions. 
In particular, there may be three solutions for $\tmax <\sqrt{3/2}$. When $|u|< | \tmax - \tmax^{-1} |$ two of the solutions, located at $|t|>\tmax$, correspond to point-like lens solutions:
\begin{equation}
t_\pm=\frac{u}{2}\left(1\pm\sqrt{1+\frac{4}{u}^2}\right)\Rightarrow \sum\left|\mu_\pm\right|=\frac{u^2+2}{u\sqrt{u^2+4}}.
\label{eq:solsunifsph}
\end{equation}
The third solution is located at $|t|<\tmax$, and does not have an analytic form, but we can determine it for $|t|\ll \tmax$, which corresponds to $u\ll|\tmax-\tmax^{-1}|$:
\begin{equation}
t_3\simeq u\left(1-\frac{3}{2\tmax^2}\right)^{-1},
\label{eq:t3unifsph}
\end{equation}
with magnification
\begin{equation}
\mu_3 \simeq \left(1-\frac{3}{2\tmax^2}\right)^{-2}\left[1-\frac{3u^2}{\tmax^2}\left(1-\frac{3}{2\tmax^2}\right)^{-2}\right]~.
\end{equation}
Note that if we take $\tmax\to0$, the magnification from this image vanishes, $\mu_3\to0$, and the total magnification is just that in~(\ref{eq:solsunifsph}). In other words, we recover the point-like magnification in Eq.~\eqref{eq:magpointlike} when we shrink the lens to the point-like limit.

Now for $\tmax<\sqrt{3/2}$ and large $u$, there is just one solution at $t = t_+$ given in Eq.~\eqref{eq:solsunifsph}. As one dials $u$ from large to small values, the number of solutions to the lens equation goes from one to three. At the transition, there are two solutions where the magnification formally diverges, since for that solution $du/dt=0$. In reality, this divergence is regulated by the finite size of the source. The existence of such caustics can, however, have an important effect, causing $u_{1.34}>1$ for the uniform sphere for some range of $r_{\rm m}/\rE$ as seen in Fig.~(\ref{fig:uT}).

For $\tmax>\sqrt{3/2}$, there is only one solution to the lensing equation regardless of $u$. For large $u$, it is $t\simeq u$ with $\mu\simeq 1$. 
For small $u$, is $t = t_3$ in Eq.~\eqref{eq:t3unifsph} with an approximate magnification of
\begin{equation}
\mu_3 \simeq 1+\frac{3}{\tmax^2}(1-u^2)~,
\end{equation}
so that to obtain $\mu > 1$ we need $u <1$, {\em i.e.} the lens must be within the point-like lensing tube.

\subsection{Power-law density profiles}
\label{app:subsec:powerlaw}
Another profile that will be useful for us is a simple power law, $\rho(r)\propto r^n$. 
For $n\ge-3$, to avoid a divergent total mass we must cut this profile off at some radius $r_{\rm m}$.
The mass profile in this case is
\beq
m(t) = \frac{\int_0^{t} d\sigma \sigma \int^{\sqrt{\tmax^2-\sigma^2}}_0 d\lambda\, \left(\sigma^2+\lambda^2\right)^{n/2}}{\int_0^{\tmax} d\gamma \gamma^{2+n}}~.
\eeq
where, again, $\tmax=r_{\rm m}/\rE$. It is useful to understand the behavior of this at small $t$. In this regime, we can simplify the expression by taking the upper limit on the $\lambda$ integral to be $\infty$, finding
\beq
m(t)\propto\left(\frac{t}{\tmax}\right)^{3+n}~.
\eeq
Thus, if $n>-2$, then $m(t)/t\to 0$ as $t\to 0$, giving $u=t=0$ as a solution to the lensing equation~(\ref{eq:lens2}). 
For a density profile that is not too diffuse, i.e. $\tmax<1$, we have $m(t=1)=1$, and so $u=0$, $t=\pm 1$ also satisfies the lensing equation. 
This means that if $n>-2$ and $\tmax<1$, there is a range of impact parameters $u$ such that there are three solutions to the lensing equation, and therefore there are impact parameters corresponding to caustics where the number of images changes abruptly and formally produces $\mu\to\infty$. 
For steeper profiles, $n<-2$, $u=t=0$ is not a solution of the lensing equation, and one thus finds only two solutions to the lensing equation, with no caustic crossings. 

We consider two power-law profiles in this work. The first is the so-called self-similar profile~\cite{Bertschinger:1985pd} 
\beq
\nn \rho (r) \propto r^{-9/4}~,
\eeq
which has been suggested to result from direct gravitational collapse of initial state perturbations within scalar condensates, such as axion miniclusters~\cite{Fairbairn:2017dmf}. 
From Sec.~\ref{app:subsec:powerlaw} of this appendix, we see that the mass profile $m(t)$ at small $t$ scales as
\beq
m(t)\propto t^{3/4}.
\eeq
Therefore, we only expect two lensed images, without any caustic crossings.
As a result we expect a smooth transition from an inefficient lens, for when the maximum radius of the subhalo is large, to a point-like lens, for when it is small. 
This is indeed the behavior we see in the critical impact parameter $u_{1.34}$ for the ``self-similar subhalo'' in Fig.~\ref{fig:uT}.

The second power-law density profile that we study is slightly shallower than the self-similar profile above,
\beq
\nn \rho (r) \propto r^{-3/2}~.
\eeq
This form is motivated by studies of the inner region of ``ultra-compact minihalos''~\cite{UCMHProfile,*Delos:2018ueo} as well as halos limited in size by free streaming~\cite{Ishiyama:2010es,*Anderhalden:2013wd,*Ishiyama:2014uoa,*Polisensky:2015eya,*Angulo:2016qof,*Ogiya:2017hbr}. 
Based on the discussion above, the mass profile at small $t$ scales as
\beq
m(t)\propto t^{3/2},
\eeq
so that $m(t)/t\to0$ as $t\to0$. 
This means that, when its size is comparable to the point-like Einstein radius, the $r^{-3/2}$ profile can give rise to caustic crossings and the associated enhancement of the microlensing magnification. 
We see this exhibited in Fig.~\ref{fig:uT} where $u_{1.34}>1$ for some values of $r_{90}/\rE$ for $\rho\propto r^{-3/2}$, similar to the uniform sphere, which can provide a reasonable approximation of this density profile for microlensing.

\subsection{NFW subhalos}
The Navarro-Frenk-White (\acro{NFW}) profile is often a good description of structures that form through hierarchical structure formation.
This profile scales as $r^{-1}$ for small radii and as $r^{-3}$ for large radii.
The scale factor $r_{\rm s}$ defines the transition between the two regimes. 
More concretely, the profile is~\cite{NFW}
\beq
\nn \rho (r) = \frac{\rho_{\rm s}}{(r/r_{\rm s})(1+r/r_{\rm s})^2}~.
\eeq
The total mass contained within this profile diverges logarithmically, so we must cut it off at some radius $r_{\rm m}$. 
As in Ref.~\cite{Fairbairn:2017dmf}, we take $r_{\rm m}=100~r_{\rm s}$ for the structures we consider, motivated by numerical studies of axion miniclusters. The mass profile $m(t)$ is obtained numerically, which we use to determine $\uT$ in Fig.~\ref{fig:uT}. Given the large ratio between $r_{\rm s}$ and $r_{\rm m}$, the \acro{NFW} profile we consider is essentially $r^{-3}$. Following the discussion above, this means that $|m(t)/t|$ is large near $t=0$, and therefore multiple images and caustics do not appear as with shallower density profiles. This means that the \acro{NFW} lens smoothly interpolates between the point-like and inefficient regimes, as seen in Fig.~\ref{fig:uT}.

\subsection{Boson stars}
Boson stars are Bose-Einstein condensates: gravitationally bound clumps of a scalar field (elementary~\cite{Ruffini:1969qy} or composite~\cite{Soni:2016gzf}) condensate, kept from collapsing under self-gravity by kinetic pressures and possibly self-repulsive forces. 
Their occupation numbers in quantum states are typically very high, hence they are described by classical field theory. It is typically sufficient to consider the non-relativistic limit,\footnote{See, however, Ref. \cite{Croon:2018ybs}.}
in which their hydrostatic equilibrium is described by Schr\"{o}dinger-Poisson equations.
In the limit of negligible self-coupling, we solve these equations numerically to compute our mass profile $m(t)$ and the resultant $\uT$ in Fig.~\ref{fig:uT}.

The Schr\"{o}dinger-Poisson equations are given by~\cite{Ruffini:1969qy}
\begin{align} \notag
    i \partial_t \psi &= - \frac{1}{2m_\phi} \nabla^2 \psi + m_\phi \Phi \psi~, \\ 
    \nabla^2 \Phi &= 4 \pi G |\psi|^2~,
\label{eq:scroedingerpoisson}
\end{align}
where $\psi$ is a non-relativistic decomposition of the scalar field $\phi$,
$$ \phi(r,t) = \frac{1}{\sqrt{2}m_\phi} e^{-i m_\phi t} \psi(r,t) + {\rm c.c.}, $$
with $m_\phi$ the mass of the scalar, and
$\Phi$ the self-gravitational potential.
The ground state of a boson star is spherically symmetric and can be written as 
\begin{equation}
    \psi_{\rm gs}(r,t) = \left(\frac{m_\phi}{\sqrt{4 \pi G}} \right) \Psi(r) e^{-i \mu t}~,
\end{equation}
where the dimensionless parameter $\Psi$ parametrizes the radial distribution, and $\mu$ is a chemical potential. 
For the ground state, the first equation in \eqref{eq:scroedingerpoisson} becomes 
\begin{align} \label{eq:bosonstareq}
    \mu \Psi = - \frac{1}{2m_\phi} \left(\Psi'' + \frac{2}{r} \Psi' \right) + m_\phi \Phi\Psi
\end{align}
in the absense of self-interactions. 
We solve these equations numerically, as no closed form solution for $\Psi(r)$ exists generally. 
As a function of radius, the enclosed mass is then given by 
\begin{equation}
    M(r) = \frac{1}{m_\phi G} \int_0^{m_\phi r} dy \ y^2 \ \Psi^2(y)~,
\end{equation}
from which the projected mass profile $m(t)$ may be computed. 
Rescalings of $\Psi$ leave Eq.~\eqref{eq:bosonstareq} invariant, hence solutions exist for any boson star mass:
\beq
\left(\frac{M}{10^{-3}~\text{M}_{\odot}}\right) \sim \lambda \left(\frac{10^{-7}~{\rm eV}}{m_\phi}\right)~,
\eeq
where $\lambda$ is the ratio of the gravitational potential energy per scalar constituent to its mass $m_\phi$~\cite{Bar:2018acw}.
For the non-relativistic treatment to be self-consistent, $0 < \lambda \ll 1$.

As seen in Fig.~\ref{fig:uT} and mentioned in Sec.~\ref{app:subsec:sphunif}, the qualitative features of boson star microlensing signals are captured by the uniform sphere toy model.
\\

\section{Speedy evaluation of event rates}
\label{app:speed}

We here outline a method to rapidly evaluate the integral in Eq.~\eqref{eq:Nevents} that, to our knowledge, has not been mentioned in the literature before. 
First we rewrite the integrand as $(A(x)/\tE^4) e^{-B(x)/\tE^2}$ to separate the $\tE$-dependent and $x$-dependent parts.
Then we note that the integral over $\tE$ has an analytic form:
\beq
\nn A \int d\tE \frac{e^{-B/\tE^2}}{\tE^4} = \frac{A}{2B} \left[ \frac{e^{-B/t^2}}{\tE} - \sqrt{\pi} \frac{{\rm erf}(\sqrt{B}/\tE)}{\sqrt{B}} \right]~.
\eeq
Using this, we can evaluate the integral over $x$ in Eq.~\eqref{eq:Nevents} for narrow bins of $\tE$, taking the efficiency $\varepsilon(\tE)$ constant in each bin. 
Summing over the bins gives $N_{\rm events}$.

\section*{Acknowledgments}

Many thanks to
Nikita Blinov,
Kfir Blum,
Joe Bramante,
David Curtin,
Jeff Dror,
Bob Holdom,
Joachim Kopp,
David Morrissey,
Ingrida Semenec,
Sarah Schon,
Jack Setford,
Sean Tulin,
Aaron Vincent,
Yue Zhao,
and especially  
Zihui Wang
and Larry Widrow,
for magnifying our work.
It is supported by the Natural Sciences and Engineering Research Council of Canada. 
T\acro{RIUMF} receives federal funding via a contribution agreement with the National Research Council Canada.

\bibliography{refs}

\end{document}

%% file: universalnewcommands.tex
\newcommand{\gsim}{\gtrsim}
\newcommand{\lsim}{\lesssim}
\newcommand{\ra}{\rightarrow}

\newcommand{\acro}[1]{\textsc{\MakeLowercase{#1}}} 
\newcommand{\osn}{\oldstylenums}
 
\newcommand{\beq}{\begin{equation}}
\newcommand{\eeq}{\end{equation}}
\newcommand{\bea}{\begin{eqnarray}}
\newcommand{\eea}{\end{eqnarray}}
\newcommand{\nn}{\nonumber}